\documentclass[11pt]{article}
\begin{document}
\sloppy
\author{\textbf{Arbab I. Arbab}
\footnote{On leave from: Comboni College for Computer Science, P.O. Box 114, Khartoum, Sudan}\\
\\
ICTP, P.O. Box 586, Trieste, Italy\\ \\ arbab@ictp.trieste.it}
\title{Conspirative cosmology with variable constants}
\maketitle \begin{abstract} We have investigated a cosmological
model with  variable speed of light $(c$), gravitational constant
($G$) and cosmological constant ($\Lambda$). The model is shown to
solve the horizon, flatness and monopole problems of the early
universe. We have found that with a certain variation of these
parameters the model predicts a cosmic acceleration. The model
also predicts that for a flat universe $\Lambda$ vanishes in both
radiation and matter phases. If the gravitational constant is
allowed to increase then we might not need the existence of dark
matter.
\end{abstract}
\vspace{2cm} \textbf{Key Words}: {Cosmology: inflation, early
universe, variable constants, cosmological parameters}
\vspace{.5in} \baselineskip=20pt
\section{Introduction}
The idea of change fundamental constant of physics was first
suggested by Dirac (Dirac, 1937, 1938). He postulated that the
gravitational constant ($G$) decreases with time ($t$) as,
$G\propto \frac{1}{t}$ . It was thought such a variation would
help understand the existence of very large numbers appearing when
one compares atomic physics with cosmology. But very recently
Moffat (1993), Albrecht and Magueijo (1999),  Barrow (1999, 2003),
Avelino \& Martins (1999) have conjectured that if the speed of
light falls at certain rate then the horizon, flatness and
monopole problems of the standard model can be solved without
recourse to inflationary paradigm. However, the variation of the
speed of light will have important consequences especially during
the nucleosynthesis era. Besides this, the variation of the speed
of light will require a critical revision of the theory of
relativity which relies upon the constancy of speed of light. The
implication of this variation will be immense in the arena of
thermodynamics, astrophysics and cosmology. This variation will
also alter our basic principle and laws in physics, e.g., Lorentz
invariance and general covariance. It seems plausible, that the
force of gravity can be integrated into the theory of
electrodynamics if the velocity of light changes according to some
cosmological law.

We will postulate here in this letter that such a change is
countered by a change in the cosmological (or gravitational)
constant, in such a way that the usual energy conservation still
holds. With this minimal change of cosmology in mind we
investigate the consequences of this variation.

\section{The model}
Solution of Einstein field equations for a universe with
cosmological constant $\Lambda$ yields the two  equations
\begin{equation}\label{1}
\left(\frac{\dot R}{R}\right)^2=\frac{8\pi
G\rho}{3}-\frac{kc^2}{R^2}+\frac{\Lambda c^2}{3}\ ,
\end{equation}
and
\begin{equation}\label{1}
\left(\frac{\ddot R}{R}\right)=-\frac{4\pi
G}{3}(\rho+\frac{3p}{c^2})+\frac{\Lambda c^2}{3} \ .
\end{equation}
The cosmic fluid is characterized by the equation of state
\begin{equation}\label{c}
p=(\gamma-1)\rho c^2\  \ , \qquad 1\le\gamma\le 2\ .
\end{equation}
  It is assumed by Barrow (2003) that the Friedmann cosmology in
the presence of time dependent speed of light $c(t)$ is described
by the same Friedmann equation as those of constant $c$.
Therefore, in what follows we will consider $c$ and $\Lambda$ as
functions of time (or scale factor). We employ the following
ansatzs for $\Lambda$ and $c$ as
\begin{equation}\label{c}
c=c_0 R^n \ ,\qquad \Lambda=\Lambda_0R^m\ ,
\end{equation}
where $n, m, c_0, \Lambda_0$ are undetermined constants. With this
prescription, eqs.(1) and (2) yield
\begin{equation}\label{1}
\left(\frac{\dot R}{R}\right)^2=\frac{kc_0^2(2-3\gamma)
}{2n+3\gamma-2}R^{2n-2}+\frac{\gamma\Lambda_0
c_0^2}{m+2n+3\gamma}R^{m+2n}+AR^{-3\gamma}\ ,
\end{equation}
where $A$ is an integration constant.
Now consider the following cases:\\
If $\Lambda\ne 0$. We see from eq.(5) that whenever the term
\begin{equation}\label{}
n< \frac{1}{2}(2-3\gamma)\ \,\ \ {\rm and}\qquad m<-2\ ,
\end{equation}
first term  in LHS of eq.(5) becomes negligibly small for large
$R$ in comparison with the other two terms. If this happens the
curvature term will be insignificantly small and the flatness
problem is automatically solved, as well as the horizon and
monopole problems. We further admit that if the universe is
dominated by strings ($\gamma=\frac{2}{3}$), then there will be no
flatness or horizon problems, as evident from eq.(5). Inflationary
solution is known to require the condition $\gamma<-\frac{2}{3}$
implying an exotic equation of state to solve these problems. We
remark here that our eqs.(5) and (6) generalize Barrow (2003)
equations to include a variable $\Lambda$ term. We arrive at
eqs.(5) and (6) using Friedmann equations directly, in comparison
with Barrow (2003) method which employs matter conservation law
[see eq.(3)].
\\
In the radiation dominated phase $\gamma=\frac{4}{3}$ so that
eq.(6) reduces to
\begin{equation}\label{}
n<-1
\end{equation}
which implies a decaying speed of light.
\section{Early acceleration}
If the universe was dominated by strings and that $\Lambda\ne 0$,
then irrespective of the value of $k$ $(i.e., 0, 1, -1)$, one
would have
\begin{equation}\label{}
\left(\frac{\dot R}{R}\right)^2=\frac{\gamma\Lambda_0
c_0^2}{m+2n+3\gamma}R^{m+2n}\ ,
\end{equation}
if $A=0$. We see that the curvature term vanishes identically. It
is evident from the above equation that, when $\Lambda_0>0$ one
has
\begin{equation}\label{}
  m+2n+2>0
\end{equation}
With this constraint eq.(8) can be easily solved to give
\begin{equation}\label{1}
R=B t^{-\frac{2}{m+2n}}
\end{equation}
where $B$= constant. The constraint in eq.(9) implies that
\begin{equation}\label{}
\ddot R >0 \ .
\end{equation}
Thus one gets a power law inflation in comparison with the
problematic inflationary solution which normally requires an existence of an unseen scalar particles. \\
Now if $A\ne 0$ but $k=0, \Lambda=0$, one has for radiation
dominated (RD) phase the equation
\begin{equation}\label{}
\left(\frac{\dot R}{R}\right)^2=AR^{-4}\ ,
\end{equation}
which can be solved to give
\begin{equation}\label{}
R=D \ t^{\frac{1}{2}}
\end{equation}
where $D=$ const. This is the usual Einstein-de Sitter solution
for RD phase.
\section{Matter dominated (MD) phase}
For matter dominated phase one has $\gamma=1$. For a flat universe
($k=0$) with $\Lambda=0, A\ne  0$, one has
\begin{equation}\label{}
\left(\frac{\dot R}{R}\right)^2=AR^{-3}\
\end{equation}
which is solved to give
\begin{equation}\label{}
R=D' t^{\frac{2}{3}}\ ,
\end{equation}
which is the usual Einstein-de Sitter solution. We observe that
this cosmology is determined by the constants $k, c, \Lambda$ and
does not depend on $G$. However, we can not here determine the
time variation of $c$ and $\Lambda$ completely.

Now if $A=0$ and $k=0, \Lambda>0$ one has the constraint
\begin{equation}\label{}
m+2n+3>0\ .
\end{equation}
Applying the above constraint in eq.(5) one obtains
\begin{equation}\label{}
R=B' t^{-\frac{2}{m+2n}}\ ,
\end{equation}
 where $B'=$ constant.
With the aid of eq.(16), eq.(17) implies that
\begin{equation}\label{}
    \ddot R >0
\end{equation}
as long as
\begin{equation}\label{0}
    m+2n>-2\ .
\end{equation}
Thus, we see that with a variable speed of light one can  solve
the horizon and flatness problems of the early universe; and
justify the present cosmic acceleration. Since our model does not
determine $m$ and $n$ uniquely, their values has to be found from
the present observational data.
\section{Conspirative mechanism (I)}
One can derive the conservation equation from eqs.(1) and (2).
This is done by differentiating eq.(1) an eliminating $\ddot R$
between eqs.(1) and (2). We treat here $c$ and $\Lambda$ as
variable but $G=$ constant. Consequently, one  obtains the
following equation
\begin{equation}\label{2}
\dot\rho+3\frac{\dot
R}{R}\left(\rho+\frac{p}{c^2}\right)=-\frac{\dot \Lambda c^2}{8\pi
G}-\frac{\Lambda c \dot c}{4\pi G}+\frac{3k c\dot c}{4\pi G R^2}
\end{equation}
We assume here that the \emph{usual} energy conservation hold so
that
\begin{equation}\label{}
\dot\rho+3\frac{\dot R}{R}\left(\rho+\frac{p}{c^2}\right)=0
\end{equation}
provided that the other scalars ($c, \Lambda$) conspire to satisfy
it, i.e.,
\begin{equation}\label{}
\frac{\dot \Lambda c^2}{8\pi G}+\frac{\Lambda c \dot c}{4\pi
G}-\frac{3k c\dot c}{4\pi G R^2}=0
\end{equation}
In this case one assumes, not  only Friedmann equation to be the
same for variable $c$ and $\Lambda$, but also the usual energy
conservation to hold too. With this assumption one can solve
eq.(22) to study the effect of the variation of $c$ on this
cosmology. Again assume the speed of light to have the form
\begin{equation}\label{}
c=c_0R^n \ ,
\end{equation}
where $c_0$ and $n$ are undetermined constants. Integrating
eq.(21) using eq.(3) to obtain
\begin{equation}\label{}
\rho=A' R^{-3\gamma}\ ,
\end{equation}
where $A'=$ constant. Integrating eq.(22) using eq.(23) one gets
\begin{equation}\label{}
\Lambda=\frac{3 k n}{(n-1)}R^{-2}+FR^{-2n}\ ,\qquad F= \rm
constant
\end{equation}
If one sets $F=0$, we get
\begin{equation}\label{}
 \Lambda=\frac{3 k n}{(n-1)}R^{-2}\ ,
\end{equation}
a variation law that already suggested by several authors
(O$\rm\ddot{z}$er \& Taha, 1986, 1987; Chen \& Wu, 1990) with
different motivation. It is interesting to link the variation of
$\Lambda$ to the variation of the speed of light. We observe that
if $n=0$ then $\Lambda=0$. This implies that if the speed of light
is constant then the cosmological constant vanishes. However, if
$k=0$ then $\Lambda=0$  whether $c$ is constant or not. Thus one
can solve the cosmological constant and flatness problems
simultaneously, in radiation and mater dominated phases. These
problems have preoccupied many scientist for long time. Some
string theorists believe that $\Lambda=0$ according to string
theory. We therefore, provide a viable model realizing this
belief. One then would attribute a non-zero cosmological constant
to a variation of $c$. Cosmologists have been seeking some
symmetry that dictates $\Lambda$ to vanish. The existence of a
non-zero $\Lambda$ in the early universe is needed for
inflationary solution. The present observations coming from type
Ia supernovae suggest that our universe is accelerating at the
present time. With the present model we have shown that this
cosmic expansion is also featured. Thus our model could give a
plausible answer to this enigma. However, we have shown that the
cosmological problems that inflation thought to solve can be
solved in the framework of the present cosmology. We come to the
conclusion that a vanishing cosmological constant follows from
either the constancy of the
speed of light and/or flatness of the universe.\\
Again, this setting solves the horizon and flatness problems in
the early universe with the same condition as the one in eq.(6).
Substituting eqs.(24) and (26) in eq.(1) one gets
\begin{equation}\label{}
\left(\frac{\dot R}{R}\right)^2=\frac{8\pi G
A'}{3}R^{-3\gamma}+\frac{(2n+1)}{(n-1)}kc_0^2R^{2n-2}\ .
\end{equation}
This time the flatness and horizon problems can be solved with
\begin{equation}\label{}
n=-\frac{1}{2}\ \  \ {\rm or} \qquad n<\frac{1}{2}(2-3\gamma) \ .
\end{equation}
For the radiation ($\gamma=\frac{4}{3}$) and matter ($\gamma=1$)
dominated phases, we obtain the usual Einstein-de Sitter
solutions, viz.,
\begin{equation}\label{}
R\propto t^{\frac{1}{2}}\ \ ,\qquad R\propto t^{\frac{2}{3}}\ \
,\qquad \Lambda=0\ .
\end{equation}
\section{Variable $c, \Lambda$ and $G$}
We will consider here, for completeness,. a model in which $c, G$
and $\Lambda$ vary with scale factor ($R$). Accordingly, eqs.(1)
and (2) yield (by differentiating eq.(1) and substituting it
eq.(2) to eliminate $R$, $\dot R$)
\begin{equation}\label{}
\rho'+\frac{3}{R}\left(\rho+\frac{p}{c^2}\right)+\rho\frac{G'}{G}-\frac{3
k c c'}{4\pi G R^2}+\frac{\Lambda' c^2}{8\pi G}+\frac{\Lambda c
c'}{4\pi G}=0\ ,
\end{equation}
where the prime $'$ denotes derivative with respect to the scale
factor ($R$). A simplest solution of eq.(30) is to consider the
power law for $c, \Lambda $ and $G$ of the form
\begin{equation}\label{}
c=c_0R^n\ ,\qquad \Lambda=\Lambda_0 R^m\ , \qquad G=G_0 R^\alpha\
.
\end{equation}
Applying eq.(31) to eq.(30) using eq.(3), one obtains
\begin{equation}\label{}
\rho=A
R^{-3\gamma-\alpha}+\frac{c_0^2(3nk-n\Lambda_0+\Lambda_0)}{4\pi
G_0(2n+3\gamma-2)}R^{2n-\alpha-2}\ ,
\end{equation}
where for consistency we take $m=-2$, and $A$= const. Substituting
eq.(32) in eq.(1) using eq.(31) we get
\begin{equation}\label{}
\left(\frac{\dot R}{R}\right)^2=\frac{8\pi G_0
A}{3}R^{-3\gamma}+\frac{c_0^2(2k-3\gamma k+\gamma
\Lambda_0)}{(2n+3\gamma-2)}R^{2n-2}\ .
\end{equation}
The horizon, flatness and cosmological constant problems will be
solved with the condition
\begin{equation}\label{}
n< \frac{1}{2}(2-3\gamma)\ ,
\end{equation}
which is the same as eqs.(6) and (28) (except here $m=-2$).
Equations (32) and (33) generalizes eqs.(\textbf{8}) and
(\textbf{9}) of Barrow \& Magueijo (1999) in their attempt to
solve the quasi-flatness and quasi-lambda problems. Thus, whether
$G$ varies or not, the solution of the horizon and flatness
problems is not affected. This solution depends only on the
variation that $\Lambda$ and $c$ assume. This is also evident as
$\alpha$ does not enter in the constraint in eq.(34), and from
eq.(33). The effect of a positive $\alpha$ is to increase the
energy density of the universe in the early times and decrease
(dilute) it at late times. This may help explain why, today, we
observe the universe to have a deficit in its anticipated energy
density (as favored by inflationary models). Hence, one need not
assume any dark matter to scale up the energy density to the
required level. A negative $\alpha$ will have the opposite
contribution. For a more realistic model of the universe, one
should consider the effect of bulk viscosity (since it is a basic
property  of any real fluid) on the evolution of the universe. In
an earlier work (Arbab, 1997), we have shown that in a viscous
universe with variable $G$ and $\Lambda$ a de-Sitter inflationary
solution arises naturally without being imposed. We will tackle
this problem in the forthcoming letter.
\subsection{Radiation and Matter dominated  phases}
We will study here the following two cases:
\subsubsection{$A=0$ \ .}
To solve eq.(33) for matter dominated phase we set $\gamma=1$, so
that for $\Lambda_0 > 0$ , one obtains
\begin{equation}\label{}
R=N_m t^{\frac{1}{(1-n)}}\ \ ,\qquad
N_m=c_0\sqrt{\frac{(\Lambda_0-k)}{(2n+1)}}\ ,\qquad n \ne\ 1\ ,
-\frac{1}{2}\ .
\end{equation}
This solution is also obtained by Barrow and Magueijo (1999) with
$G$ constant [see eq.(15)]. \\
To solve eq.(33) for radiation dominated phase we  set
$\gamma=\frac{4}{3}$, so that that for $\Lambda_0 > 0$ , one
obtains
\begin{equation}\label{}
R=N_r t^{\frac{1}{(1-n)}}\ \ ,\qquad
N_r=c_0\sqrt{\frac{(2\Lambda_0-3k)}{3(n+1)}}\ ,\qquad n \ne\ 1\ ,
-1\ .
\end{equation}
We  see that the scale factor $R$ has the same form but $n$
assumes different values in the two eras (RD \& MD). This really
shows that the way $c$ varies affect the expansion of the universe
that result in solving may of the cosmological problems.
\subsubsection{$A\ne 0$\ .}
In this case we will assume that
\begin{equation}\label{}
(2k-3\gamma k+\gamma \Lambda_0)=0\ \ , \ {\rm and} \qquad
2n+3\gamma-2\ne 0 \ ,
\end{equation}
in both eras (RD \& MD). This is guaranteed if one takes $k=0$ and
$\Lambda_0=0$. For this situation the solution of eq.(33) reduces
to the Einstein -de Sitter solution, viz.,
\begin{equation}\label{}
R\propto t^{\frac{1}{2}} \ \  \ ({\rm RD})  \ \ {\rm and}  \qquad
R\propto t^{\frac{2}{3}}\ \ \ ({\rm MD}) .
\end{equation}
\section{Conspirative mechanism (II)}
We assume again that the \emph{usual} energy conservation
\begin{equation}\label{}
\dot\rho+3\frac{\dot R}{R}\left(\rho+\frac{p}{c^2}\right)=0
\end{equation}
to hold. Integrating eq.(39) one obtains
\begin{equation}\label{}
\rho=R^{-3\gamma}\  \, \qquad A=\rm const.
\end{equation}
Using eq.(39) in eq.(30) one gets
\begin{equation}\label{}
\rho\frac{G'}{G}-\frac{3 k c c'}{4\pi G R^2}+\frac{\Lambda'
c^2}{8\pi G}+\frac{\Lambda c c'}{4\pi G}=0\ ,
\end{equation}
As before we adopt the power law for $c$ and $\Lambda$ of the form
\begin{equation}\label{}
c=c_0 R^n\ \ , \qquad \Lambda=\Lambda_0 R^m\ .
\end{equation}
Using eqs.(42) and (40), eq.(41) can be integrated to give
\begin{equation}\label{}
G=\frac{c_0^2}{4\pi
A}\frac{(3nk-n\Lambda_0+\Lambda_0)}{(2n+3\gamma-2)}R^{2n-2}+B\ ,
 \qquad m=-2\ , \qquad B=\rm const.
\end{equation}
Substituting eqs.(40), (42) and (43) in eq.(1) yields
\begin{equation}\label{6}
\left(\frac{\dot R}{R}\right)^2=\frac{c_0^2(\gamma\Lambda_0
-3k\gamma+2k)}{(2n+3\gamma-2)}R^{2n-2}+DR^{-3\gamma} \ ,\qquad
D=\frac{8\pi AB}{3}\ .
\end{equation}
We observe that
\begin{equation}\label{}
\Lambda\propto R^{-2}\ ,
\end{equation}
a variation that is favored by several cosmologists (e.g., Chen \&
Wu, O$\rm\ddot{z}$er \& Taha).
\subsection{Radiation and matter dominated phases}
We study here two cases as in \textbf{6.1}:\\
\subsubsection{D=0 \ .}
In this case eq.(44) reads
\begin{equation}\label{6}
\left(\frac{\dot R}{R}\right)^2=\frac{c_0^2(\gamma\Lambda_0
-3k\gamma+2k)}{(2n+3\gamma-2)}R^{2n-2}\ .
\end{equation}
This can be integrated to give
\begin{equation}\label{}
R=N_m t^{\frac{1}{(1-n)}}\ \ ,\qquad
N_m=c_0\sqrt{\frac{(\Lambda_0-k)}{(2n+1)}}\ ,\qquad n \ne\ 1\ ,
-\frac{1}{2}\ .
\end{equation}
during  matter dominated phase, and
\begin{equation}\label{}
R=N_r t^{\frac{1}{(1-n)}}\ \ ,\qquad
N_r=c_0\sqrt{\frac{(2\Lambda_0-3k)}{3(n+1)}}\ ,\qquad n \ne\ 1\ ,
-1\ ,
\end{equation}
during  radiation dominated phase. We notice that eqs.(35) and
(36) are identical with eqs.(46) and (47), respectively. This
shows that our assumption of eq.(39) was correct. We have already
applied a similar conservation law for a bulk model (Arbab, 1997).
\subsubsection{$D\ne 0$\ .}
In this case we will assume that
\begin{equation}\label{}
(2k-3\gamma k+\gamma \Lambda_0)=0\ \ , \ {\rm and} \qquad
2n+3\gamma-2\ne 0 \ ,
\end{equation}
in both eras (RD \& MD). This can be satisfied if one takes $k=0$
and $\Lambda_0=0$. For this situation the solution of eq.(33)
reduces to the Einstein -de Sitter solution, viz.,
\begin{equation}\label{}
R\propto t^{\frac{1}{2}} \ \  \ ({\rm RD})  \ \ {\rm and}  \qquad
R\propto t^{\frac{2}{3}}\ \ \ ({\rm MD}) .
\end{equation}
Which is a solution we have already obtained in eq.(38). Thus, in
an evolving universe particles  interact (conspire) in such a way
the energy conservation law holds. Hence, energy conservation is
indeed a stringent law that governs the dynamic of the interacting
particles in our expanding universe.
\section{Avelino \& Martins cosmology}
Avelino \& Martins (1999) propose a generalization of general
relativity that includes the variation of $c$ and $G$. They
however, argued that the this theory is both covariant and Lorentz
invariant. This is parameterized by writing
\begin{equation}\label{}
\frac{G}{c^2}=\rm const\equiv A\ .
\end{equation}
We apply this equation in eq.(20) with the variation
\begin{equation}
 c=c_0 R^n\ .
 \end{equation}
 We then obtain
\begin{equation}
\rho=\frac{3nk}{4\pi
A}\frac{1}{(2n+3\gamma-2)}R^{-2}+MR^{-2n-3\gamma}\ ,\qquad M=\rm
const.,
\end{equation}
where we have taken for simplicity $\Lambda=0$. \\
Now substitute eqs.(51)- (53) in (1) to get
\begin{equation}\label{6}
\left(\frac{\dot
R}{R}\right)^2=\frac{kc_0^2(2-3\gamma)}{(2n+3\gamma-2)}R^{2n-2}+D'R^{-3\gamma}\
,\qquad D'=\frac{8\pi c_0^2A M}{3}
\end{equation}
It is clear that this solves the flatness and horizon problems if
\begin{equation}\label{}
n<\frac{1}{2}(2-3\gamma)\ ,
\end{equation}
which is the same as eq.(34). However. Avelino and Martins claim
that the horizon and flatness problems require similar conditions
to the one found in the context of the standard cosmological
model. They argued that a theory that reduces to General
Relativity in appropriate limit and solves the horizon and
flatness problems of the standard model must either violate the
strong energy condition or Lorentz invariance, or covariance. They
also show that the variation in eq.(51) guarantees conservation of
mass energy density $\rho$. The solution of eq.(54) is already
worked out in sec. \textbf{5, 6, 7}. We notice that Barrow (1999,
2003) has obtained many  solutions that we have found in this
letter by different methods. We conclude that Avelino \& Martins
cosmology is equivalent to Albrecht and Magueijo, Barrow, and
Moffat models.
\section{Conclusion}
We have studied here the effect of the variation of speed of light
and cosmological constant in cosmology. We have found that such a
variation can in principle solve many of the long persisting
problems of standard model of cosmology. Of these problems are the
flatness, horizon, monopole and cosmological constant problems. A
conspiracy between these variable constants can still preserve
some of the standard model picture. The variation of the
gravitational constant is not dealt with here, but will be one of
our future endeavors. These models share some features of the
original Moffat, Albrecht and Barrow scenario. Our model provides
a good solution to the existence of dark matter in the universe if
the gravitational constant varies with time. It also helps
understand the cause of the present cosmic acceleration. Moreover,
the cosmological constant is found to relax as $\Lambda\propto
R^{-2}$ which is suggested by some cosmologists.
\section*{Acknowledgements}
I would like to thank the abdus salam International Center for
Theoretical Physics (ICTP) for hospitality and Comboni College for
providing financial support.
\section*{References}
Albrecht, A., and Magueijo, J., {\it Phys. Rev.}, \textbf{D59}, 043516 (1999).\\
Arbab, A.I., {\it Gen. Rel. Gravit.}, \textbf{29}, 61 (1997).\\
Avelino, P. P., and Martin, C.J.A.P., {\it Phys. Lett.} , \textbf{B459}, 468 (1999).\\
Barrow, J.D., {\it Phys. Lett. }, \textbf{B564}, 1 (2003).\\
Barrow, J.D., and Magueijo, J., {\it Phys. Lett. }, \textbf{B447}, 246 (1999).\\
Barrow, J.D., {\it Phys. Rev.}, \textbf{D59},04515 (1990).\\
Chen, W., and  Wu, Y., {\it Phys. Rev.} \textbf{D 41}, 695 (1990).\\
Dirac, P. A. M., {\it Proc. R. Soc.}, \textbf{A165}, 199 (1938).\\
Dirac, P. A. M., {\it Nature} \textbf{139}, 323 (1937).\\
Magueijo, J., {\it Phys. Rev.}, \textbf{D62},103521 (2000).\\
Moffat, J. W., {\it Int. J. Mod. Phys.}, \textbf{D2},351 (1993).\\
O$\rm\ddot{z}$er, M., and  Taha, M. O.,  {\it Phys.Lett.},\textbf{171B}, 363 (1986).\\
O$\rm\ddot{z}$er, M., and  Taha, M. O., {\it Nucl. Phys.} \textbf{B287}, 776 (1987).\\
\end{document}